# The SPACE THEA Project

**Martin Spathelf and Oliver Bendel**

School of Business FHNW, Bahnhofstrasse 6, CH-5210 Windisch
martin.spathelf@gmail.com; oliver.bendel@fhnw.ch

### Abstract

In some situations, no professional human contact can be available. Accordingly, one remains alone with one's problems and fears. A manned Mars flight is certainly such a situation. A voice assistant that shows empathy and assists the astronauts could be a solution. In the SPACE THEA project, a prototype with such capabilities was developed using Google Assistant and Dialogflow Essentials. The voice assistant has a personality based on characteristics such as functional intelligence, sincerity, creativity, and emotional intelligence. It proves itself in seven different scenarios designed to represent the daily lives of astronauts, addressing operational crises and human problems. The paper describes the seven scenarios in detail, and lists technical and conceptual foundations of the voice assistant. Finally, the most important results are stated and the chapters are summarized.

## Introduction

Advances in space technology in recent years have greatly increased the likelihood of a manned flight to Mars. However, since this will take several months, it is important that the astronauts receive the best possible support on their journey, both professionally – for example, for repairs – and psychologically.

Today's voice assistants can already "understand" (acoustically perceive and classify) what a user says very well. In many cases, they can also perform the correct action or give the correct answer when prompted. Prominent examples of such voice assistants are Google Assistant, Siri from Apple, or Alexa from Amazon.

Voice assistant Clarissa and social robot CIMON were developed for space travel. Clarissa's task was to guide "an astronaut through potable water analysis procedures" (NASA 2005). CIMON (version 2) "is used to perform routine tasks, such as documenting experiments, searching for objects and taking inventory, as well as explaining complex information and instructions regarding scientific experiments and repairs to the vehicle" (Martin and Freeland 2021).

To date, however, no satisfactory voice assistant with empathic skills has been developed for a Mars flight. Accordingly, Oliver Bendel carried out a project at the School of Business FHNW in 2021 to close this gap. By then, he and his teams had gained a lot of experience with chatbots that recognize problems of the user (Bendel et al. 2017; Bendel 2018). These were mainly developed to target issues in machine ethics and social robotics.

The artificial woman SPACE THEA – the acronym stands for "The Empathic Assistant for Space" – is intended to be a contribution to American space travel (Spathelf 2021). She is supposed to be able to recognize emotions to some extent, respond to the astronaut in selected scenarios, and, above all, display empathy (and certain emotions). Ultimately, this will increase the astronauts' ability to work and their well-being. The prototype fits within the discipline of social robotics – according to Bendel (2021), conversational agents such as chatbots and voice assistants may also be counted as social robots if they exhibit certain characteristics. Machine ethics (Anderson and Anderson 2011) plays a sideline role in the project.

## Conditions of the SPACE THEA Project

The goal of the SPACE THEA project was to create an empathic voice assistant for a Mars flight (Spathelf 2021). The authors established the following parameters and approaches derived from the goal:

1. Find acceptance among users: This point is addressed indirectly in the project. The team tries to increase theoretical acceptance. The effectiveness of the implementation is not empirically proven.
2. Acoustically understand what the user has said and give an acoustic answer to the question asked: This is done by an already existing voice assistant framework. The optimization of this process is thus limited to the reliability of the framework.

3. Use lifelike artificial voice: The artificial voice is also provided by a voice assistant framework. For this reason, the voice is dependent on the choice of voice assistants. The voice should sound genuine and trustworthy as well as be able to be influenced, for example in terms of tone, pitch, softness, and euphoria.
4. Recognize emotions and respond accordingly: This point depends heavily on the flexibility of the voice assistant framework. Deriving emotions from the content of what is spoken takes high priority.
5. Understand the user's intention and provide a response based on it: To be able to respond empathically to the different situations of the astronauts, the voice assistant must also be trained to respond to such situations. In the project, various scenarios are used in which it has to hold its own.

This formulated the most important requirements in the project. Some of them are directly related to technical issues. Therefore, these will be explained first. Then relationship and personality aspects are discussed.

## Technical Implementation

In the project, various frameworks for voice assistants were compared and evaluated (Spathelf 2021). In particular, the focus here was on requirements 2, 3, and 4 from the previous section. Among the candidates were Google Assistant and Dialogflow Essentials (Google 2021a/b), Google Assistant and Rasa, and Alexa from Amazon.

Google Assistant and Dialogflow Essentials met the most requirements and offered full integration of a customizable voice assistant (Spathelf 2021). There are many different voices. The framework offers enough freedom for development. Dialogflow is a system from a subsidiary of Google and handles dialogue processing. Google Assistant is responsible for text-to-speech and speech-to-text. Google Assistant and Dialogflow was the best choice for this project.

The Google Assistant Voice offers voices for the concrete voice assistant. In the project, Female 2 (EN-CA) was chosen, a voice with a Canadian accent. This comes from the subjective taste and perception of the team. In practice, the gender and type of voice should be free to select.

The original plan was to use Speech Synthesis Markup Language (SSML) to make the voice as pleasant and expressive as possible. However, the chosen tools do not allow many possibilities here. Nevertheless, some SSML modifications were achieved.

The voice assistant is available in a secure area. First, you log in to the corresponding Google account. Then you call up a subpage of https://dialogflow.cloud.google.com. The "integrations" link takes you to the test area, where you can communicate with the voice assistant. A conversation can be listened to via https://youtu.be/Ij---G1TgSY.

## User's Relationship to the Voice Assistant

Developing a voice assistant that is accepted by the user as a point of contact in various situations (requirement 1) involves some difficulties (Spathelf 2021). The establishment of a relationship is necessary for humans to achieve the required emotional intimacy in certain situations. This has particular significance in the case of very personal issues, e.g., when the user is struggling psychologically. In such a situation, a certain amount of trust is required from the person for him or her to open up.

According to Reis and Shaver, intimacy is an exchange process in which personal thoughts and feelings are revealed to a counterpart. If the other person reacts positively to what is expressed, there is a greater chance that the relationship between the two people will be strengthened as a result (Reis and Shaver 1988, p. 375). According to Laurenceau et al. (1998), intimacy develops over repeated interactions over time and is important for the user to open up emotionally. With each interaction, a perception is formed that reflects the level of intimacy and the meaning of the relationship.

According to these explanations, intimacy is an important component of any relationship. For the user to perceive the communication as engaging, the voice assistant should try to understand, accept, and validate the user in factual and emotional contexts. According to Laurenceau et al. (1998), in any interaction, perceived qualities and individual differences can influence the user's behavior. If the perceived motives and needs differ strongly from the interests of the counterpart, this can have a negative influence.

If the relationship between the user and the voice assistant is to be strengthened, it is also important that the mutual exchange of personally relevant information or emotions takes place (Laurenceau et al. 1998). This poses a challenge because social robots do not have an actual ability to feel and suffer (Bendel 2021). When the user says something, it can be interpreted and responses can be made accordingly, but the voice assistant does not feel anything in the process. Thus, the reactions are not really based on consciousness, feelings, or motivations, but are artificially generated (Poushneh 2021) – it is nothing but a simulation.

This problem cannot be completely circumvented at the present time. However, it is still important for the user to be able to establish a personal relationship with the voice assistant for some of the envisioned scenarios to work. As mentioned earlier, this would be difficult to achieve if it were to operate exclusively on a factual level. So there are two conflicting sides: On the one hand, a relationship must be established with the help of intimacy; on the other hand, social robots (if you want to count SPACE THEA among them) have no internally motivated expressions of feelings.

According to Poushneh (2021), although humans can most likely rationally determine that a voice assistant is displaying fabricated emotions, studies show that they can still be influenced by them. Apart from the ethical aspects, which should be considered, this finding gives some freedom to build a relationship with the user despite the voice assistant's lack of intrinsic motivation. Ultimately, the effects of anthropomorphizing the voice assistant will be positive.

## Personality of SPACE THEA

The problem of lacking intrinsic motivation can be further minimized by integrating a kind of personality into the voice assistant (Spathelf 2021). This promotes the user's impression that he or she is talking to a consistent and predictable counterpart (which in turn is related to anthropomorphism). In the case of SPACE THEA, her personality is expressed through her statements and voice. According to Tamm and Serena (2011), the positive influence was found to be at least increased in Europeans and Americans. In Asian culture, this trait did not bring any significant advantage.

Since the prototype is intended for American space travel, this approach makes most sense. In addition, the internal motivation can now be mapped indirectly with the help of this personality. In concrete terms, this means that the motivations and goals of the voice assistant are included in every conversation. The personality flows into the dialogue processing with every extension of the voice assistant. Ultimately, however, it should only serve as a guideline in dialogue creation and not as an inescapable law.

It is important to establish some basic principles before fleshing out the personality. When the user talks to the voice assistant, they should be able to forget to some extent that they are not talking to a human. On the other hand, the voice assistant should have enough "self-awareness" to "understand" that it may not be viewed by all humans as such. Moreover, it may itself point out its machine-ness (and associated inadequacy) (Bendel 2018). Dialogue creation should thus consider what the user's perspective might be in relation to the voice assistant in any given situation.

According to Tolmeijer et al. (2021), there is still insufficient empirical evidence as to which gender is best received by the user. Nevertheless, some case reports from companies suggest that female voices are preferred over male voices. Of course, the validity of case reports is to be doubted and the scientific validity is accordingly small. Nevertheless, a gender has to be chosen (if one does not access a neutral synthetic voice), and with SPACE THEA a decision was made to use a female voice.

The voice assistant sees itself as female within the simulation and presents itself accordingly. This fact is reflected in the voice in the end. It will also introduce itself with the name SPACE THEA. Although it sees itself as female, it brings enough "self-awareness" to this that a human might not actually perceive it as a female entity.

An important aspect that should be considered when developing the personality is how the user should ideally perceive the voice assistant. Its ultimate goal is to help the astronaut in various situations and to be a good companion. However, this can only be achieved if the user is satisfied with the voice assistant and actually uses it.

According to Poushneh (2021), these aspects can be influenced by increasing the perceived control and confidence during interactions with the voice assistant using various personality traits. In her study, approximately 50 personality traits were measured and classified into seven categories, namely functional intelligence, aesthetic attraction, protective qualities, sincerity, creativity, sociability, and emotional intelligence. The study was limited to Microsoft's Cortana, Google Assistant, and Amazon's Alexa. They are functionally oriented voice assistants. The characteristics with the best influence on user behavior were functional intelligence, sincerity, and creativity (Poushneh 2021).

SPACE THEA is additionally used for building a relationship and empathic interaction with the user. For this reason, emotional intelligence will most likely also be important with her. During the dialogue elaboration, the focus was accordingly on the use of functional intelligence, sincerity, creativity, and emotional intelligence.

- According to Poushneh (2021), creativity reflects how effective the voice assistant is at providing information. In short, it is about how trendy, smooth, and original the voice assistant feels to the user.
- Functional intelligence refers to the degree of effectiveness, efficiency, reliability, and usefulness of the information obtained using the voice assistant. Accordingly, it describes the practical benefit that one has if one continues to listen to it (Poushneh 2021).
- According to Poushneh (2021), emotional intelligence refers to the speech assistant's ability to be perceived as human. It also describes how empathetic, humorous, and humble it is – and thus how much it is a conversational partner who responds to the emotional needs of the other person.
- According to Poushneh (2021), sincerity shows how honest, sympathetic, original, friendly, down-to-earth, and appealing the voice assistant's information is. The user recognizes whether it has their best interests at heart and whether it can be trusted.

It is difficult to integrate sincerity into the personality. The reason for this lies in a problem already mentioned. For example, if the voice assistant says, "I'm sorry.", this would in effect be lying, since it doesn't feel anything – so it's not really sorry. This is true if sincerity is viewed from the perspective of an inanimate digital assistant. If it is viewed from the perspective of the people behind the development of a voice assistant, this picture may change. The emotions that the voice assistant explicitly expresses are, in the end, bestowed by humans. This means that if they put themselves in the personality of it during development and act from its point of view, these emotions can potentially be just as honestly meant as if they came from a real person (Spathelf 2021).

## Including Machine Morality

The personality of the voice assistant can be supplemented – machine ethics is responsible for this – with a machine morality (Anderson and Anderson 2011; Bendel 2019). This is then integrated into the dialogue. The voice assistant itself, as mentioned earlier, has no actual internal motivations or feelings, no consciousness, and no free will. The proper moral approach is to ensure that it is implemented in the best interest of the user. No harm should come from SPACE THEA having an absent or incorrect moral principle. The voice assistant should be able to empathically respond to the user (Spathelf 2021).

What proves to be a challenge in personality design is the lack of control over the situations in which the voice assistant finds itself (Spathelf 2021). Personality, as mentioned earlier, can only be implemented indirectly. According to Bendel (2019), certain situations can be anticipated, but they may turn out differently than expected. Although morality remains the same, its relevance may change depending on the situation. Thus, if the situation changes, it may be necessary to also consider the applied morality from a different perspective than an immobile machine could (Bendel 2019).

For example, with a voice assistant like SPACE THEA, which cannot recognize vocal information but only content information, it could go like this: A user tells it in a sad tone of voice that he or she is doing well, and it takes this literally and responds accordingly with pleasure. The voice assistant, even with the right moral compass, would respond correctly according to the information given. However, in this case, it would not have all the relevant information because it cannot hear the sad tone. If it empathized and tried to understand the person as well as the problem, this would be a better reaction. Although it would be clear in advance that this case could occur, there may be no way in this project to counteract it (with the help of voice recognition you could implement something like that though). In the prototype, such compromises are accepted and dialogues are adjusted to predefined scenarios.

Regardless of the situation, it is arguable what is the morally correct response in a complex situation, as morality, while possibly shared by many, is still intrinsically subjective to the voice assistant. For the development of a machine morality, it is necessary to determine what should be implanted in the personality of the voice assistant. The personality, in combination with the situation, ultimately determines how the voice assistant responds to the user.

An interesting moral approach in this context is utilitarianism, a form of consequentialism, for which the fundamental principle is the maximization of total utility. This states that the morally right thing to do is to maximize the sum of the welfare of all those affected by an action (Bendel 2019). Since the voice assistant is not a sentient person, it is pushed out of the affected group. This means it acts for the benefit of the entire human crew on the spaceship (like in the example of a conversation flow, see Tab. 1), but excludes itself from the equation. This ensures that it does not pursue actions (mediated via the programmer) that are an end in themselves.

## Scenarios of SPACE THEA

Since the creation of a complete voice assistant is not possible due to time constraints, the project is limited to a few scenarios (Spathelf 2021). These are intended to depict different situations that could occur on the journey to Mars and take into account requirement 5. The scenarios are all intended to show a different emotional state of the user and demonstrate the reactions of the voice assistant. They were found through team brainstorming. In practice, subject matter experts from the space community should identify and prioritize relevant situations for a Mars flight.

### Overview of the Scenarios

The following scenarios are exemplary for the prototype:

a) Technical and operational support of the astronaut – neutral situation
b) Crisis situation on the spacecraft – tense and hectic situation
c) Waking up/Greeting in the morning – everyday situation
d) Insulting the voice assistant – testing the limits of the voice assistant
e) An astronaut is not doing so well – stressful situation
f) Interview with the voice assistant – bringing people and the voice assistant closer together
g) General dialogues – maintaining the impression of a real interlocutor

There are three questions which should be asked in each scenario (Spathelf 2021):

1. What is the situation the user is experiencing?
2. How might the user feel in this situation?
3. How should SPACE THEA respond based on the answers to these two questions?

All of these questions will be addressed below.

**Details of the Scenarios**

In this section, the authors elaborate on each of the scenarios (Spathelf 2021). They state the goal and outline of each of them. Then they list additional relevant information. Where possible, the appropriate personality traits are also outlined. For the sake of economy, an example of a conversation process can only be included for one of the situations.

a) Technical and operational support of the astronaut

The goal of this scenario is to represent a common situation. The voice assistant adopts a pragmatic, emotional mood and is helpful to the user in an ordinary situation.

The scenario looks like this: An astronaut is looking for the electric switch in one of the rooms on the spacecraft. SPACE THEA is supposed to help find it.
– SPACE THEA should be able to give different answers based on variable voice input. For example, if the user asks for the switch in the bathroom, they will get a different answer than if they ask for the switch in the cockpit.
– In this scenario, functional intelligence is needed.

b) Crisis situation on the spacecraft

The goal of this scenario is to depict a crisis or stress situation. An astronaut then primarily needs a voice that tries to calm him or her down. In addition, the voice assistant should help to find a solution to the problem in a short time.

The scenario is to take place during an engine problem. The astronaut loses his or her nerve, and the voice assistant accompanies him or her through the crisis.
– SPACE THEA should act solution-oriented and at the same time be aware that the astronaut is tense and could make mistakes faster.
– It is assumed that she is aware of the wrong actions in this situation. However, it is not certain which is the correct action. SPACE THEA has no decision-making authority in this situation, that is, she may not make a decision in place of the crew.
– Above all, functional intelligence in combination with emotional intelligence is needed.

Here is an example of a conversation flow (Tab. 1):

| |
|---|
| User – Report the problem: "We have a problem with the engine! I don't, I don't, I don't know what to do." |
| Regardless of how much information SPACE THEA has about the state of the spacecraft, it notices from this statement that something is wrong. From the phrase "I don't, I don't, I don't", it can be inferred that the user may be stressed and overwhelmed. Now it is important to both empathically talk to the user and pragmatically guide them to solve the problem. |
| SPACE THEA – Attempt to calm down and motivate: "I understand. Let me help you. What would be the best option? Let's keep calm and think this through together." |
| SPACE THEA does not take the initiative to solve the problem, nor does she pass judgment on the situation. She only tries to reassure the user as much as possible and encourage him or her to solve the problem. |
| User – Proposed solution: "I really have no idea. Shutting down engine 1 might be an option. And power the others." |
| This answer is only meant to serve as an example and has no relation to a scenario that could occur on a real spaceship. Nevertheless, in this situation it is assumed that this would be the correct solution. |
| SPACE THEA – Encourage further steps: "That sounds like a very sound plan. Maybe check in with the crew if possible and then try it, but we have to act fast." |
| If the user's idea does not sound completely outlandish, SPACE THEA encourages further steps. If there is still time, she still advises him or her to exchange ideas with the crew, if he or she has not already done so. |

Tab. 1: Conversation flow according to (Spathelf 2021)

Fig.1 shows the schematic representation of the flow of the conversation:

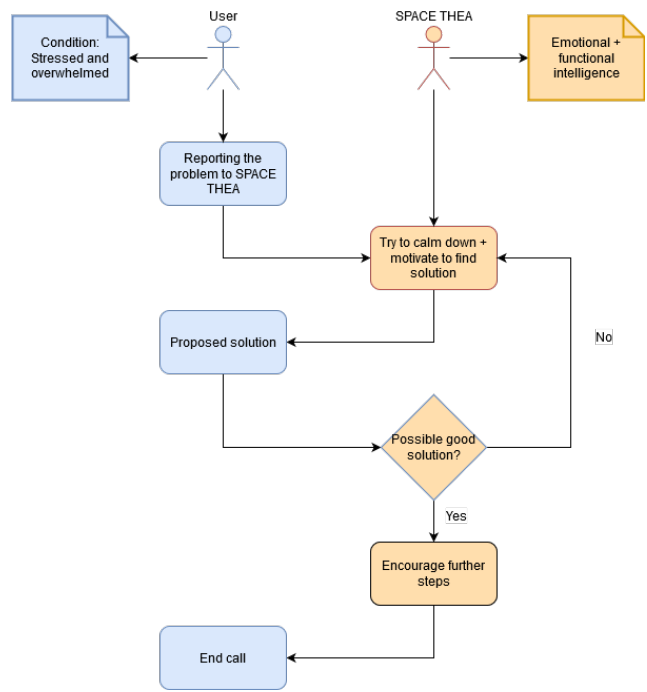

Fig. 1: Schematic representation according to (Spathelf 2021)

c) Waking up/Greeting in the morning

The goal of this scenario is to depict a situation that an astronaut might encounter in everyday life. He or she should be greeted in the morning to facilitate the start of the day. In this situation, the voice assistant should respond to the astronaut's needs in a friendly and considerate manner.

The scenario takes place directly in the morning or after sleeping. The astronaut has not slept well and SPACE THEA responds to this situation. This situation is to be used to further strengthen the relationship in a simple, everyday way.

– For practical use, an event would be needed to notice the astronaut waking up. For example, the dialogue could be triggered by the ringing of the alarm clock. Due to the lack of an event, this is done with the words "I woke up".
– The precondition is that SPACE THEA already knows the name of the user. This is necessary so that it can greet him or her with the appropriate name. If the name is still unknown, Dialogflow cannot execute this scenario.
– Both creativity and emotional intelligence are used.

d) Insulting the voice assistant

The goal of this scenario is to test the limits of the voice assistant. It should let the user know that insults are not welcome, but in a way that shows understanding. SPACE THEA is not intended to annoy the user while still trying to maintain mutual respect.

The scenario looks like this: The astronaut is frustrated because something did not work as it should. He or she then starts communicating with SPACE THEA and insults her without her having done anything wrong.

– This scenario is only used when SPACE THEA is offended without a reason directly related to her.
– She makes the user understand that she does not like gratuitous insults. At first glance, it looks like she is acting from a moral point of view with this action as an end in itself. However, the goal here is not to save face, but to uphold mutual respect, so that the intimacy with the user can be maintained or even improved.
– It is intended to use mainly emotional intelligence, but also sincerity. The latter is considered from the point of view of SPACE THEA's personality. For example, when she says "I do not appreciate that.", she is speaking according to her personality.

e) An astronaut is not doing so well

The goal of this scenario is a short therapy. The voice assistant tries to address a psychological problem with an astronaut. In doing so, it should show consideration and understanding for his or her problem and help him or her to feel better.

The scenario envisions an astronaut who is plagued by loneliness and would like to see friends and family on Earth again. He or she should feel heard and understood. It is not a matter of solving the user's problem, but of standing by them and giving them as much support as possible. SPACE THEA assumes the role of a therapist and a friend at the same time.

– Parts of this scenario were inspired by an interview with Leaf and Nelson (Mackenzie 2020). They say that such a conversation can be broken down into several steps. First, the feelings should be validated. Second, questions should be asked to encourage the user to self-reflect. Third, it is about reassuring the users that SPACE THEA is there for them and should ask if there is anything she can do for them.
– Emotional and functional intelligence as well as sincerity are used.

f) Interview with the voice assistant

The goal of this scenario is to establish a relationship between the voice assistant and the user. Above all, he or she should gain insights into the "thinking" and "feeling" of the voice assistant right from the start.

The scenario is supposed to represent the getting-to-know phase between the astronaut and SPACE THEA. He or she tries to get closer to her with the help of some questions. She answers from the programmed feeling and tries to leave the best possible impression.

– The conversation flow in this scenario follows a structure that also appears in a short video called "Detroit: Become Human | Chloe | PS4" (PlayStation 2018). This helps cover the most important points when getting to know each other. The questions lend themselves to giving surprising answers.
– Both creativity and sincerity are shown.

g) General dialogues

The goal of the last scenario is for the voice assistant to be able to conduct general dialogues. These should help to better maintain the impression of an existence similar to ours.

SPACE THEA can draw on a repertoire of questions and answer options. So, if the user asks something like "Are we friends?", she should be able to answer that. General questions and statements from the user are already covered by Google Assistant and still need to be enriched with personality-specific answers.

– The general dialogues include only one-sentence answers. Thus, no structured dialogues result from it, but SPACE THEA is only able to answer simple questions or statements.
– Several personality traits are possible.

Of course, other scenarios can be formed. But even if you double the number, it is far from covering every situation on the long journey.

# Results of the Project

In this section, the main results and findings of the SPACE THEA project are discussed (Spathelf 2021). The technical and theoretical perspectives are outlined and combined.

## Technical Aspects

From the user's point of view, a microphone is required in the periphery of the voice assistant into which he or she can speak. Furthermore, a voice is necessary that gives him or her an answer to a question via a speaker. For a question from the user to be answered, the voice assistant relies on a background system to process the information correctly.

First, what is spoken must be converted into text so that the voice assistant can decode the information. The text can then be assigned to a scenario or user intent. Each intention found contains a response, which is then output to the user again via the speaker using text-to-speech. The voice assistant also needs a way to incorporate dynamic objects into conversations.

Often, the user's statements must be understood within a context. Conversation trees are necessary to describe these so that the voice assistant can be expanded. They were used to map the scenarios and the more complex conversations.

## Theoretical Aspects

To develop an empathic voice assistant, it was important to consider the user's perception. It is often significant for a person to have a positive connection or relationship with another person before opening up. This relationship aspect was taken into consideration when creating the dialogue so that the users can be primed to reveal their feelings. Their feelings should be both recorded and validated in the dialogues so that he or she can walk away from the conversations feeling as positive as possible. Basically, in each conversation, as stated, three questions were considered.

Another important aspect was how the user should ideally perceive SPACE THEA. The joy of interaction and satisfaction with the voice assistant should be increased as much as possible. Therefore, the integration of a personality was seen as a possibility to create a structure from which an inner motivation could be simulated. In this project, functional and emotional intelligence, creativity, and sincerity were identified as the most useful personality traits. In addition, the moral principle of maximizing the overall benefit of the spacecraft's passengers was incorporated into the personality. This should help to ensure that the voice assistant acts in the best interest of the entire crew in every conversation.

## Combination of Technical and Theoretical Aspects

What aspects must now be considered when combining the technical and theoretical with each other in practice (the practice of SPACE THEA and in any future implementation)?

The speech-to-text engine of Google Assistant is relatively good, but there are cases where the user is still not understood. This could be counterproductive for an empathic voice assistant. For example, if the user says "I'm feeling lonely." and the voice assistant misunderstands him or her, he or she may not say it again and a very important conversation will be lost.

Voice customization should serve to make a voice assistant sound more natural. However, Google Assistant is far from perfect in this regard. Especially the modification with SSML is, as mentioned, only possible to a certain extent. In practice, the customized voice often fails to achieve its desired effect and can only be used in a limited way.

During implementation, a conflict of goals may arise between generalizing and concretizing the conversation. For example, if the user utters the phrase "What is your name?" or "Hi, what is your name?" both would most likely be assigned to the same scenario. In principle, this is also desirable under most circumstances. In the case of the second statement, SPACE THEA should greet you back with a "Hi", but this should never happen in the first example. This knowledge leads to the fact that the answers must be generalized to a certain degree, so that no wrong answers are given. Unfortunately, however, important statements by the user that the voice assistant could have addressed are possibly lost in a more generalized process.

It is very difficult to implement the complexity of a situation in its entirety in conversation trees. In the end, the voice assistant is only as good as this complexity can be represented. As long as the user follows the intended path of the conversation, he or she can be provided with a good experience. However, if he or she deviates too far from it or makes innuendos that the voice assistant does not understand, the experience will be worse because his or her artificial counterpart can no longer follow the conversation.

Since the person that the voice assistant is conversing with is unknown, it cannot be specifically addressed. At most, the role, that of the astronaut, can be considered. It has been outlined how the exchange process between user and voice assistant shapes the relationship. However, since everyone expects a slightly different interaction, it is clear that the dialogues chosen will not be specific to the user with whom SPACE THEA will ultimately converse in practice.

## Testing of SPACE THEA

Several tests by the team have proven that the voice assistant can hold its own in the scenarios, especially if you know the questions it can answer. However, technical difficulties arose once due to Google's framework, and SPACE THEA had to be set up again. It was not evaluated with test groups how well the voice assistant works with free input.

## Summary and Outlook

The main goal of the project was to create an empathic voice assistant – ideally more empathic than Clarissa and CIMON – that responds adequately to a user's emotions and situations for operation on a Mars flight. The authors looked at the development from both a technical and theoretical perspective. Subsequently, the results from this research were combined and implemented in a voice assistant using predefined scenarios.

To capture the user's interest, the voice assistant must establish a personal relationship. It must be possible to trust it. Likewise, one must be able to assume that one will not be deceived if one opens up. This paper described what such relationship building might look like. In the future, the assumptions and statements would need to be adjusted with additional empirical evidence.

To give the voice assistant some consistency, it has been assigned a form of personality. Of course, this involves more than just a few personality traits, a moral rationale, and a male, female, or neutral voice. The personality of a real human being is complex and consists of many facets. Therefore, it remains open whether, first, a more complex personality can be represented and, second, whether the complete personality can be projected onto a practical situation.

Further, ethical considerations were made for the development of the voice assistant and the maximization of the total benefit – in this case the spaceship's team – was determined as a basic moral principle. This has the advantage of being universally applicable. Of course, this is not the only existing moral principle, and perhaps it can be replaced by a better one.

Google Assistant and Dialogflow provide a powerful platform and tool kit. Nevertheless, there were some problems that could only have been solved with extensive effort. It remains to be seen how technology will progress and to what extent it can be used to implement even better empathic voice assistants. There are some details, especially for such assistants, that would be important to make the dynamic between human and machine as positive as possible. Some of them have been mentioned in this paper, and it is likely that more such difficulties will show up in the design of a larger and more complex voice assistant.

The project focused on astronauts, but the approach also has the potential to be applied to other people or situations. For example, the voice assistant could help a person who is suffering from loneliness and communication problems. Such a direction would enlarge the effectiveness of the proposed approach.

The developed voice assistant SPACE THEA is a prototype and therefore has limited practical applicability. All in all, however, it can be said that the project has achieved the defined goals. In the tailored scenarios, the voice assistant responds empathically and competently where necessary, and overall, convincingly. It thus contributes to the professional and personal well-being of the imagined astronaut. The project is only a small step of a small team – the use of a successor of SPACE THEA on a spacecraft, however, would be a big step for mankind.